\documentclass[conference,a4paper]{IEEEtran}
\usepackage[left=1.43cm,right=1.43cm,top=1.82cm,bottom=4.3cm]{geometry}

\usepackage[algo2e]{algorithm2e} 
\usepackage{amsmath,amssymb,amsmath,amsfonts}
\usepackage{bm}
\usepackage{bbm}
\usepackage{graphicx}
\usepackage{cite}
\usepackage{epstopdf}
\usepackage{gensymb}
\usepackage{color}
\usepackage{lipsum}
\usepackage{float}
\usepackage{epstopdf}
\usepackage[mathcal]{eucal}
\usepackage{subfiles}
\usepackage[T1]{fontenc}
\usepackage{siunitx}
\usepackage{tabulary}
\usepackage{array}
\usepackage{adjustbox}
\usepackage[dvipsnames]{xcolor}
\usepackage{cite}
\usepackage{times}
\usepackage{placeins}
\usepackage{textcomp}
\usepackage[font=small]{caption}
\usepackage{flushend}
\usepackage{color, colortbl}
\usepackage{tabularx}
\usepackage{bm}
\usepackage{enumerate}
\usepackage{tikz}
\usepackage{pgfplots}
\usepackage{epstopdf}
\usetikzlibrary{shapes,arrows}
\usepackage[utf8]{inputenc}
\usepackage{mathtools}
\usepackage[utf8]{inputenc}
\usepackage{xcolor}
\usepackage{tikz}
\usetikzlibrary{chains,arrows,calc,positioning}
\usepackage{amssymb}
\usepackage{pifont}
\usepackage{soul}
\usepackage{breqn} 
\usepackage{booktabs} 
\usepackage{multirow}
\usepackage{enumitem}
\usepackage{tabulary}
\usepackage[normalem]{ulem}
\usepackage{dblfloatfix}
\usepackage{makecell}
\usepackage{lipsum}
\usepackage{amsthm}
\usepackage{comment}
\usepackage{filecontents}
\usepackage{subcaption} 

\usepackage[nolist,nohyperlinks]{acronym}

\newtheoremstyle{mystyle}
  {}
  {}
  {\itshape}
  {}
  {\bfseries}
  {.}
  { }
  {}
\theoremstyle{mystyle}

\newlength \figwidth
\definecolor{bittersweet}{rgb}{1.0, 0.44, 0.37}
\definecolor{glaucous}{rgb}{0.38, 0.51, 0.71}
\definecolor{gainsboro}{rgb}{0.86, 0.86, 0.86}
\definecolor{babyblueeyes}{rgb}{0.63, 0.79, 0.95}
\definecolor{silver}{rgb}{0.75, 0.75, 0.75}
\definecolor{neoncarrot}{rgb}{1.0, 0.64, 0.26}
\definecolor{Gray}{gray}{0.9}
\definecolor{LightCyan}{rgb}{0.88,1,1}
\definecolor{BackgroundLightBlue}{rgb}{0.97,0.97,1}
\definecolor{BackgroundGray}{gray}{0.98}

\newcommand{\red}[1]{{\textcolor[rgb]{1,0,0}{#1}}}

\newcommand{\nicola}[1]{\textcolor{magenta}{[Nicola: #1]}}

\makeatletter

 \let\oldforeign@language\foreign@language
 \DeclareRobustCommand{\foreign@language}[1]{%
   \lowercase{\oldforeign@language{#1}}}

\def\nb0{{\mathbf{0}}}
\def\nb1{{\mathbf{1}}}

\def\ncalC{{\mathcal{C}}}

\def\ncalR{{\mathcal{R}}}

\def\ncalU{{\mathcal{U}}}

\makeatother

\IEEEoverridecommandlockouts
\IEEEoverridecommandlockouts\IEEEpubid{\makebox[\columnwidth]{ \hfill} \hspace{\columnsep}\makebox[\columnwidth]{ }}
\begin{document}

\bstctlcite{IEEEexample:BSTcontrol}

\title{Capacity and Power Consumption of Multi-Layer\\6G Networks Using the Upper Mid-Band\vspace{-0.5cm}}

\author{\IEEEauthorblockN{
David L\'{o}pez-P\'{e}rez$^{\sharp}$,\vspace{0.1cm} 
Nicola Piovesan$^{\flat}$, and
Giovanni Geraci$^{\dagger\,\star}$}
\\ \vspace{-0.3cm}
\normalsize\IEEEauthorblockA{
\makebox[0.5\textwidth][r]{$^{\sharp}$\emph{Univ. Politècnica de València, Spain}} \enspace \enspace
\makebox[0.5\textwidth][l]{$^{\flat}$\emph{Huawei Technologies, France}} \\
\makebox[0.5\textwidth][r]{$^{\star}$\emph{Univ. Pompeu Fabra, Barcelona, Spain}} \enspace \enspace
\makebox[0.5\textwidth][l]{$^{\dagger}$\emph{Telefónica Scientific Research, Spain}}
}
}

\maketitle

\begin{abstract}

This paper presents a new system model to evaluate the capacity and power consumption of multi-layer 6G networks utilising the upper mid-band (FR3). The model captures heterogeneous 4G, 5G, and 6G deployments, analyzing their performance under different deployment strategies.
Our results show that strategic 6G deployments, non-co-located with existing 5G sites, significantly enhance throughput, with median and peak user rates of 300\,Mbps and exceeding 1\,Gbps, respectively.
We also emphasize the importance of priority-based cell reselection and beam configuration to fully leverage 6G capabilities. While 6G implementation increases power consumption by 33\%, non-co-located deployments strike a balance between performance and power consumption. 

\end{abstract}

\section{Introduction}


We stand on the cusp of a new era---the \ac{6G} of mobile communication technology. 
The \ac{ITU} has already outlined the requirements for \ac{6G}~\cite{ITUM2160}, 
and \ac{3GPP} 
recently agreed on a development timeline~\cite{6Gtimeline}. 
Once again, exciting times lie ahead.

To gain a perspective on \ac{6G}, 
let us first reflect on the key advancements of the \ac{5G}.
From our viewpoint, \ac{5G} adoption has been built on three main pillars~\cite{dahlman20235g}: 
\begin{itemize}
    \item 
    A key software development: 
    \emph{network slicing}.
    \item 
    An appealing technology: \emph{\ac{URLLC}}, which has seen limited success due to Wi-Fi's dominance in indoor use cases~\cite{GalGerCar2024,OugGerPol2024}.
    \item
    A hardware and signal processing breakthrough: 
    \emph{\ac{mMIMO}}.
\end{itemize}
Other features specified in \ac{5G}, while relevant and promising, are yet to gain widespread market traction. A notable example is \ac{mmWave} technology, whose spectrum licenses did not sell as expected globally, indicating a lack of interest from operators in this power-hungry technology.

We predict that 6G development and market adoption will follow a path similar to \ac{5G}. While many foresee a vast array of new functionalities, we anticipate three key components: a software breakthrough, a compelling new use case, and a hardware evolution:
\begin{itemize}
    \item 
    Integrating \emph{\ac{AI}} will be central across all protocol stack levels and control systems~\cite{8755300}.
    \item
    \emph{Joint communication and sensing} will likely serve as the key marketing driver for 6G\cite{9705498}. However, similar to \ac{URLLC}, we remain cautious about its market adoption, given the strong position of incumbents in this segment.
    \item 
    Hardware advancements will include breakthroughs in system-on-chip technology for \ac{AI} processing and further advancements in \ac{mMIMO}, leading to \emph{extreme \ac{mMIMO}} operating in the upper mid-band (7--24\,GHz), referred to as \ac{FR3} in \ac{3GPP}\cite{10459211}. 
\end{itemize}

Commercial 6G will center around \ac{FR3} and advancements in \ac{mMIMO}. The introduction of multiple 400\,MHz channels per operator in the upper mid-band offers significant potential to support demanding \ac{6G} use cases, providing an ideal balance between coverage and capacity for both outdoor and indoor deployments.
The success of this band will depend on advanced \ac{mMIMO} technology, with up to 256 \acp{TRX} and many more passive antennas required to achieve median speeds of up to 1\,Gbps per user. 
However, further developing \ac{mMIMO} as we did for \ac{5G} may result in prohibitive power consumption, and it remains unclear whether co-located 6G deployments alongside existing 5G sites will be effective.

In this paper, we evaluate the capacity and power consumption of multi-layer 6G networks using FR3. Our analysis highlights the superior performance of strategic 6G deployments non-co-located with 5G, achieving median user rates of up to 300 Mbps and peak rates exceeding 1\,Gbps. We also emphasize the critical role of priority-based cell reselection and beam codebook configuration in fully leveraging the capabilities of 6G. While 6G implementation results in a 33\% increase in power consumption, non-co-located deployments offer the best trade-off between performance gains and power consumption. Our findings underscore the need for a targeted approach to 6G deployment, paving the way for sustainable next-generation networks.

%
%
\section{Related Work on Cellular Deployments in FR3}
\label{sec:literature}

In this section, we summarize the key insights from the most relevant studies on \ac{FR3} operation.


\subsubsection*{Spectrum Allocation}

\ac{3GPP} Release 16 initiated the exploration of \ac{NR} in the 7–24\,GHz range, 
establishing a regulatory framework shaped by \ac{WRC} and \ac{ITU} outputs~\cite{3GPP38820}. 
The spectrum potentially allocated for mobile services on a primary basis amounts to approximately 9\,GHz, 
significantly enhancing capacity and complementing the increasingly congested sub-6\,GHz bands. 
Coexistence with incumbent services, 
particularly in the 12\,GHz band used by \ac{LEO} constellations like Starlink, 
poses a key challenge in \ac{FR3}~\cite{cui20246gwirelesscommunications724}.
Advanced beamforming strategies that leverage ephemeris data can mitigate interference to satellite \ac{UL}, 
while maintaining terrestrial performance~\cite{kang2024terrestrialsatellitespectrumsharingupper}.

\subsubsection*{Channel Propagation and MIMO Technology}

As noted in~\cite{cui20246gwirelesscommunications724}, 
\ac{FR3} systems experience significantly less path loss than \ac{mmWave} bands,
with minimal rain attenuation below 10\,GHz, 
making the 7--10\,GHz range resilient to adverse weather. 
Compared to 3.5\,GHz,
path losses at 10\,GHz increase by 9.5\,dB in \ac{LoS} and 9.7\,dB in \ac{NLoS}.
\ac{FR3} also has lower indoor penetration loss than \ac{mmWave}, 
making it a better option for outdoor-to-indoor coverage.
At 10\,GHz, 
penetration losses are comparable to 3.5\,GHz, 
ensuring strong connectivity from outdoor \acp{BS} to indoor environments~\cite{10459211}.
Advances in \ac{mMIMO} are crucial for fully exploiting \ac{FR3}. 
Building on the foundational 4G \ac{MIMO} systems with up to 32 \acp{TRX}, 
5G introduced \ac{mMIMO} with configurations of up to 64 \acp{TRX} and 128 antenna elements, 
extending to 384 elements in later implementations~\cite{Nokia2023extremeMassiveMIMO}.
For 6G, \ac{mMIMO} configurations with up to 256 \acp{TRX} are expected.
These advancements can significantly boost both \ac{DL} and \ac{UL} coverage, 
mitigating above losses and maintaining stable performance~\cite{Huawei2024cmWave}.
They will also increase spatial multiplexing capabilities, leading to higher data rates~\cite{Huawei2024cmWave,BjoKarKol2024}. 

\subsubsection*{Power Consumption}

While \ac{NR} is more energy-efficient than \ac{LTE}, 
power consumption remains a concern,
with current \ac{NR} deployments consuming up to three times more power than \ac{LTE}~\cite{Huawei2020}, 
and radios accounting for 73\% of total network power consumption~\cite{GSMA20205Genergy}. 
The transition to 6G, 
which could involve a significantly higher number of \acp{TRX}, 
may require innovative solutions such as hybrid beamforming, energy-saving modes, and reduced analog-to-digital conversion resolution to manage power effectively and ensure sustainability~\cite{Nokia2023extremeMassiveMIMO}.

Currently, no studies evaluate the performance and power consumption of multi-layer 4G, 5G, and 6G networks using \ac{FR3}. 
Modern mobile networks are heterogeneous, multi-layer, and multi-technology, 
with varying capabilities across layers. 
Existing \ac{3GPP} models, 
which are based on single-layer deployments and single-frequency operation, 
do not fully capture the complexities of these networks in terms of performance and power consumption.

\section{System Model for Multi-Layer 6G Networks}
\label{sec:system_model}

We now present a system model that builds upon existing \ac{3GPP} 4G and 5G system-level models to analyze the capacity and power consumption of practical 6G networks. These networks will consist of multiple layers, operate at different frequencies, and incorporate various technologies.


\subsection{User Distribution and Multi-Layer Network Deployment}

In the following, we outline the user distribution and multi-layer network configurations used to analyze the capacity and power consumption of 6G deployments.

\subsubsection*{User Distribution}

\begin{figure}[t]
    \centering
    \includegraphics[width=0.45\textwidth]{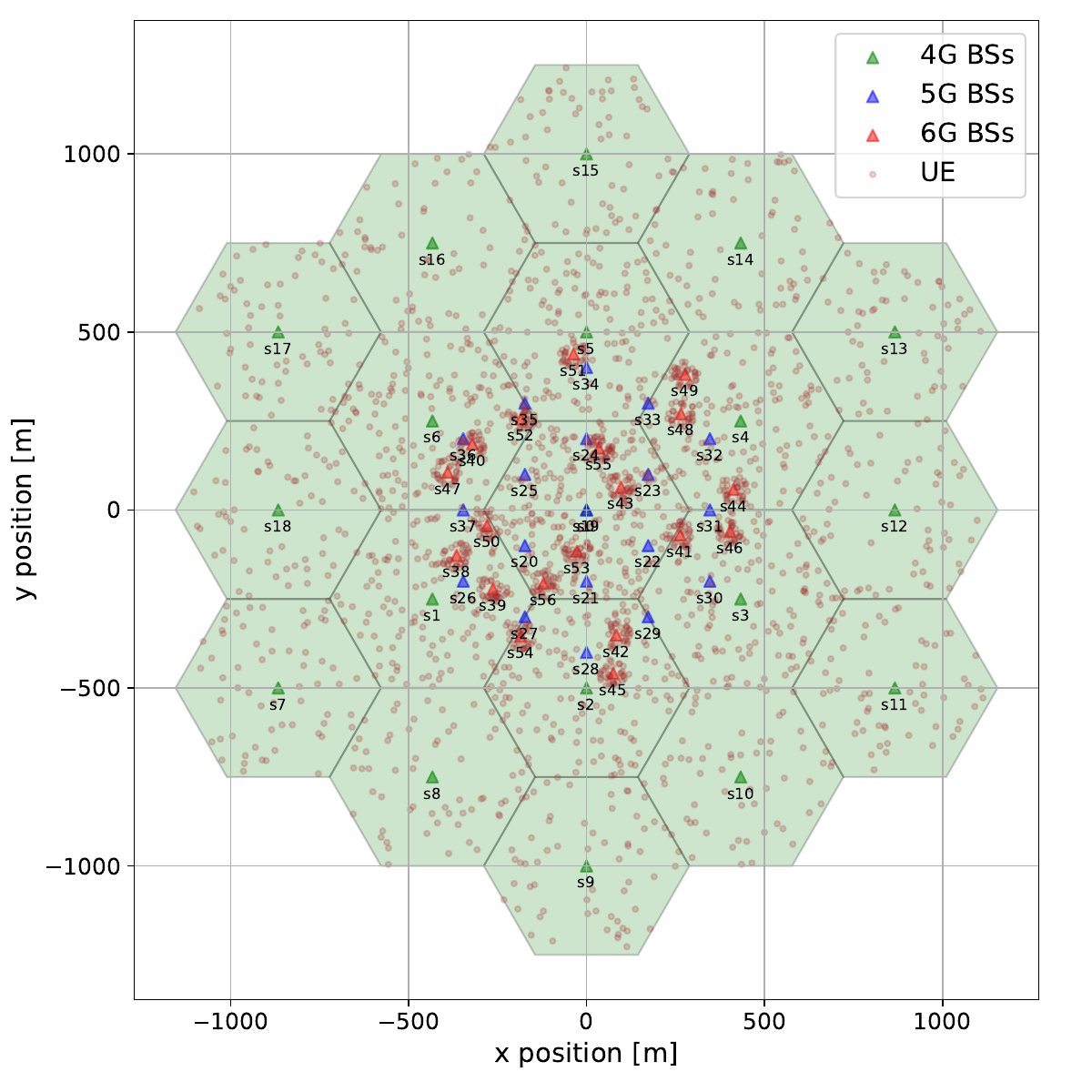}
    \caption{Inhomogeneous UE distribution served by a non-co-located 4G/5G/6G deployment, with 6G deployed at the hotspots.}
    \label{fig:scenario}
\end{figure}

The set of \ac{UE} is denoted as $\ncalU$, and \ac{UE} features (e.g., locations, height, indoor/outdoor) follow the \ac{UMa} deployment model, with an \ac{ISD} of 500\,m, as specified in \cite{3GPP38901}. To reflect nonhomogeneous traffic across regions, the number of \acp{UE} per geographical macro area (hexagon) varies. Specifically, the central hexagon hosts 80 \acp{UE} per sector, while the first and second tiers host 40 and 20 \acp{UE} per sector, respectively.
Additionally, we introduce 19 hotspots, uniformly distributed within a 500\,m radius from the scenario's center, each containing 30 \acp{UE} deployed within a 40\,m radius. Hotspots are at least 80\,m apart from one another. Fig.~\ref{fig:scenario} illustrates this deployment. Each \ac{UE} is equipped with two antennas.

\subsubsection*{Network Layers}

The network may consist of up to three layers to serve the \acp{UE}: 
\begin{itemize}
    \item 
    A 4G layer operating at 2\,GHz.
    \item     
    A 5G layer operating at 3.5\,GHz.
    \item 
    A 6G layer operating in the \ac{FR3} band at 10\,GHz.
\end{itemize}

Each layer follows one of the following deployment strategies, all based on a \ac{3GPP} two-tier hexagonal grid:
\begin{itemize}
    \item 
    A \ac{UMa} deployment with an \ac{ISD} of 500\,m.
    \item 
    An \ac{UMi} deployment with an \ac{ISD} of 200\,m.
    \item 
    A hotspot deployment with a site at the centre of each \ac{UE} hotspot.
\end{itemize}

\subsubsection*{Radio Units}

To simulate realistic deployments and assess power consumption, 
we model both co-located and non-co-located setups using a variety of radio units. The seven types of radios considered are:
\begin{enumerate}
    \item
    4G-only macro radios: 20\,MHz, 8\,\acp{TRX}, 46\,dBm.
    \item    
    5G-only macro radios: 100\,MHz, 64\,\acp{TRX}, 49\,dBm.
    \item
    5G-only micro radios: 100\,MHz, 64\,\acp{TRX}, 44\,dBm.
    \item
    6G-only micro radios: 400\,MHz, 128\,\acp{TRX}, 44\,dBm.
    \item
    6G-only pico radios: 400\,MHz, 128\,\acp{TRX}, 41\,dBm.
    \item
    Multiband 4G/5G macro radios: Combining 1) and 2).
    \item
    Multiband 5G/6G micro radios: Combining 3) and 4).
\end{enumerate}

Each single-technology radio powers three cells, covering 120º to form a tri-sector site, while multiband radios power six cells---three per technology. Details on each radio type configuration are provided in Table~\ref{tab:radio_characteristics}.

\begin{table*}[h]
\centering
\begin{tabular}{|l||c|c|c|c|c|c|c|}
\hline
\textbf{Characteristic} & \textbf{4G Macro} & \textbf{5G Macro} & \textbf{5G Micro} & \textbf{6G Micro} & \textbf{6G Pico} & \textbf{4G/5G Macro} & \textbf{5G/6G Micro} \\ \hline \hline

\textbf{Carrier frequency [GHz]}                          & 2.0 & 3.5 & 3.5 & 10 & 10 & 2.0/3.5 & 3.5/10 \\ \hline

\textbf{Bandwidth [MHz]}                              & 20 & 100 & 100 & 200 & 200 & 20/100 & 100/200 \\ \hline

\textbf{PRBs}                                         & 100 & 273 & 273 & 273 & 273 & 100/273 & 273/273 \\ \hline

\textbf{Supported cells}                              & 3 & 3 & 3 & 3 & 3 & 6 & 6 \\ \hline

\textbf{\acp{TRX} (\(M^{\rm{TRX}}_{\rm{av}}\))}       & 8 & 64 & {64} & 128 & 128 & 8/64 & 64/128 \\ \hline

\textbf{Antenna elements ($M$)}                       & 8 & 64 & {64} & 128 & 128 & 8/64 & 64/128 \\ \hline

\textbf{SSB beams (\({|\bf W}^{\rm ssb}|\))}          & 4 & 8 & 8 & 16 & 16 & 4/8 & 8/16 \\ \hline

\textbf{CSI-RS beams (\({|\bf W}^{\rm csi-rs}|\))}    & 8 & 64 & {32} & 128 & 128 & 8/64 & 64/128 \\ \hline


\textbf{TX power per Cell (\(P_{\mathrm{TX}}\)) [dBm]}& {46} & {49} & {44} & {44} & {41}& {46/49} & {44/44}\\ \hline






\end{tabular}
\caption{Characteristics of different radios and their power consumption parameters.}
\label{tab:radio_characteristics}
\end{table*}   

\subsubsection*{MIMO Arrays}

Each cell transmits using a planar \ac{MIMO} array with \(M = 2 \cdot M^{\rm h} \cdot M^{\rm v}\) cross-polarized antenna elements arranged in \(M^{\rm h}\) horizontal columns and \(M^{\rm v}\) vertical rows: 
\begin{itemize}
\item 
4G cells: \(M^{\rm h} = 2 \times M^{\rm v} = 2\), \(M = 8\). 
\item 
5G cells: \(M^{\rm h} = 8 \times M^{\rm v} = 4\), \(M = 64\).
\item 
6G cells: \(M^{\rm h} = 16 \times M^{\rm v} = 4\), \(M = 128\).
\end{itemize}
Cross-polarized antenna elements are spaced at half-wavelength intervals.

\subsubsection*{Propagation Channel}

We adopt the \ac{3GPP} \ac{UMa} and \ac{UMi} statistical channel models from \cite{3GPP38901} for the \ac{UMa} and \ac{UMi} deployments, respectively, with hotspot deployments following the \ac{UMi} variant.
For each \ac{UE} $u \in \mathcal{U}$ and cell $c \in \mathcal{C}$, 
the large-scale channel is defined as: 
\begin{equation} \label{eq:LargeScaleGain}
    \beta_{u,c} = \rho_{u,c} \; \tau_{u,c} \; g_{u,c},
\end{equation}
where $\rho_{u,c}$ represents the path loss (a function of the \ac{LoS} probability), $\tau_{u,c}$ is the spatially correlated shadow fading, and $g_{u,c}$ is the antenna gain.
Small-scale fading is modeled as Rician-distributed, producing the complex channel vector ${\bf h}_{u,c}$ between the antenna of \ac{UE} $u$ and antenna element $m$ of cell $c$.
The two channels produced by the cross-polarized antenna elements are orthogonal, with a 90-degree phase shift.

\subsubsection*{Network Configurations}

Using the definitions above, we benchmark the performance of the following seven deployment strategies, where \texttt{[$\cdot$]} indicates co-location:
\begin{enumerate}
    \item 
    \texttt{4G\,UMa}: A single-layer 4G network with a \ac{UMa} deployment using 19 4G-only macro radios.
    \item
    \texttt{5G\,UMa}: A single-layer 5G network with a \ac{UMa} deployment using 19 5G-only macro radios.
    \item
    \texttt{[4G\,UMa + 5G\,UMa]}: A two-layer co-located 4G/5G network with \ac{UMa} deployment using 19 multiband 4G/5G macro radios.
    \item
    \texttt{4G\,UMa + 5G\,UMi (UMa BS)}: A two-layer non-co-located 4G/5G network with a 4G \ac{UMa} deployment using 19 4G macro radios and a 5G \ac{UMi} deployment using 19 5G macro radios.
    \item
    \texttt{4G\,UMa + 5G\,UMi}: A two-layer non-co-located 4G/5G network, 
    with a 4G \ac{UMa} deployment using 19 4G macro radios and a 5G \ac{UMi} deployment using 19 5G micro radios.
    \item
    \texttt{4G\,UMa + [5G\,UMi + 6G\,UMi]}: A three-layer 4G/5G/6G network, 
    with a 4G \ac{UMa} deployment using 19 4G macro radios and a 5G/6G co-located \ac{UMi} deployment using 19 multiband 5G/6G micro radios.
    \item
    \texttt{4G\,UMa + 5G\,UMi + 6G\,HS}: A three-layer non-co-located 4G/5G/6G network, with a 4G \ac{UMa} deployment using 19 4G macro radios, a 5G \ac{UMi} deployment using 19 5G micro radios, and a 6G hotspot deployment using 19 6G pico radios (see Fig. \ref{fig:scenario}).
\end{enumerate}

\noindent The resulting sets of radios and cells from these deployments are denoted as $\ncalR$ and $\ncalC$, respectively.


    


\subsection{Network Operation}

We now detail the beam configuration, cell association and reselection, and data transmission phase.

\subsubsection*{SSB Beam Codebook}

In 4G networks, cell selection for \acp{UE} is based on sectorized transmitted \acp{CRS}, which do not use beamforming. 
In contrast, 5G---and presumably 6G---networks rely on beamformed \ac{SSB} beams to mitigate the higher path loss in higher frequency bands. 
In 5G and 6G cells, each \ac{SSB} beam \(s\) is generated using a complex codeword \({\bf w}_{s,c}^{\rm ssb}\), selected from a predefined \ac{SSB} codebook \({\bf W_c}^{\rm ssb}\). 
In our analysis, the \ac{SSB} codebook is generated using \ac{2D-DFT} precoding. 
Only one polarization panel is used to generate \ac{SSB} beams.
The number of \ac{SSB} beams \({|\bf W_c}^{\rm ssb}|\) depends on the number of \acp{TRX} \(M^{\rm{TRX}}_{\rm{av},c}\) and antenna elements $M_c$, 
and is detailed in Table~\ref{tab:radio_characteristics}.

\subsubsection*{Cell Association}

Each \ac{UE} $u$ identifies its serving cell by measuring the \ac{RSRP} ${\rm rsrp}^{\rm ssb}_{u, s, c}$ from each \ac{SSB} beam $s$ of each cell $c$, calculated as follows:
\begin{equation}
\label{eq:rsrpSSBComputation}
    {\rm rsrp}^{\rm ssb}_{u, s, c} = \beta_{u,c} \; \left| {\bf h}^{\rm dl}_{u,c} \, {\bf w}_{s,c}^{\rm ssb} \right|^2 \; p_{s,c}^{\rm ssb}, 
    \end{equation}
where $p_{s,b}^{\rm ssb}$ is the transmit power allocated by cell $c$ to \ac{SSB} beam $s$. 
The \ac{UE} $u$ selects its serving \ac{SSB} beam $\hat{s}_u$ and serving cell $\hat{c}_u$ by choosing the one that maximizes the measured \ac{RSRP}.

\subsubsection*{Priority-Based Cell Reselection}

After the initial \ac{UE} association, cell reselection ensures that \acp{UE} connect to the most suitable network layer, even if the cell providing that layer is not the strongest, provided the necessary channel conditions are met.
Reselection updates the serving \ac{SSB} beam $\hat{s}_u$ and cell $\hat{c}_u$ using a priority-based system. The strongest beam that meets the minimum signal strength thresholds and corresponds to the highest priority is selected. In our model, 4G is assigned Priority 0, 5G Priority 1, and 6G Priority 2, with signal thresholds of $-110$\,dBm for 5G and $-108$\,dBm for 6G. Further optimization of these thresholds on a per-cell basis is left for future work.

\subsubsection*{Data Transmission Phase}

\begin{figure*}[!htb] 
\normalsize
\setcounter{equation}{7}
\scriptsize

\begin{equation} 
 \label{eq:SINR_Computation}
\gamma_{u,l,k} = 
    \frac{
    \beta_{u, \hat{c}_u} \; \left| {\bf h}^{\rm dl}_{u, \hat{c}_u, k} \, {\bf w}^{{\rm dl}, l}_{u, \hat{c}_u, k} \right|^2   p^{{\rm dl}, l}_{u,\hat{c}_u,k}
    }
    {
    \sum\limits_{p \in \mathcal{U}_{\hat{c}_u} \setminus u}
    \beta_{u, \hat{c}_u}  
    \left| {\bf h}^{\rm dl}_{u, \hat{c}_u, k} \, {\bf w}^{{\rm dl}, l}_{p, \hat{c}_u, k} \right|^2   p^{{\rm dl}, l}_{p,\hat{c}_u,k} + 
    \sum\limits_{c \in C \setminus  \hat{c}_u} 
    \sum\limits_{i \in \mathcal{U}_{c}}
    \beta_{u, c}
    \left| {\bf h}^{\rm dl}_{u, c,k} \, {\bf w}^{{\rm dl}, l}_{i, c,k}\right|^2  p^{{\rm dl}, l}_{i,c,k}  
    +   
    \sigma^2_k
    }  
\end{equation}

\normalsize
\setcounter{equation}{8}
\hrulefill
\vspace*{4pt}
\end{figure*}

\begin{figure}[t]
    \centering
    \includegraphics[width=0.5\textwidth]{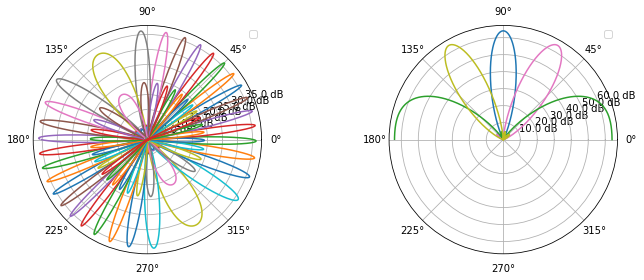}
    \caption{Horizontal (left) and vertical (right) diagrams of the 64 beams generated using a \ac{2D-DFT} codebook with \(M^{\rm h} = 16 \times M^{\rm v} = 4\) antenna elements in a single polarization. 
    Horizontal and vertical boresight angles are set to 30º and 90º, respectively.}
    \label{fig:beams}
\end{figure}

To exploit beamforming and multiplexing in 4G, 5G, and 6G, we adopt a Type I \ac{CSI}-based operational approach \cite{3GPP38214}. 
Each cell \(c\) transmits \acp{CSI-RS} for channel estimation, using a \ac{2D-DFT} codebook \({\bf W_c}^{\rm csi-rs}\), similar to the \ac{SSB} codebook \({\bf W_c}^{\rm ssb}\) but with a larger number of \acp{CSI-RS} beams to provide finer spatial resolution (see Table~\ref{tab:radio_characteristics} and Fig. \ref{fig:beams}).
Both polarization panels are used to generate the \ac{CSI-RS} beams, resulting in two fully orthogonal beams per direction. 
Each \ac{UE} \(u\) measures the \ac{RSRP} ${\rm rsrp}^{\rm csi-rs}_{u, s\prime, \hat{c}_u}$ for each \ac{CSI-RS} beam $s\prime$ of its serving cell $\hat{c}_u$ and reports the index of the strongest beam $\hat{s\prime}_u$. 
Based on this, the serving cell $\hat{c}_u$ selects a specific precoding codeword \({{\bf w}^{\rm dl \nearrow}_{u, \hat{c}_u} = \bf w}_{\hat{s\prime}_u,\hat{c}_u}^{\rm csi-rs}\) for data transmission to \ac{UE} \(u\).
Given that \acp{UE} have two antennas, the complementary beam \({{\bf w}^{\rm dl \searrow}_{u, \hat{c}_u}}\) in the orthogonal polarization is also used to enhance the data rate. 
For simplicity, we refer to the allocated \ac{MIMO} beams/layers assigned to a \ac{UE} as \(l \in \{1,L\}\), with \(L=2\) in this case.

\subsection{Key Performance Indicators}

We now define the metrics used to evaluate the capacity and power consumption of multi-layer 6G deployments.

\subsubsection*{SINR and Effective SINR}

The data \ac{SINR} $\gamma_{u,l,k}$ at \ac{UE} $u$ in \ac{MIMO} layer $l$ and \acp{PRB} $k$ is computed using \eqref{eq:SINR_Computation}.
Here, 
$\mathcal{U}_c$ denotes the subset of \acp{UE} associated with cell $\hat{c}_u$---the serving cell of \ac{UE} $u$---, while 
${\bf h}_{u,c,k}^{\rm dl}$ is the downlink channel gain from \ac{UE} $u$ to cell $c$ in \acp{PRB} $k$, and
${\bf w}_{u,c,k}^{{\rm dl}, l}$ and $p_{u,c,k}^{{\rm dl}, l}$ are the allocated precoding codeword and transmit power for \ac{UE} $u$ by cell $c$ on \ac{MIMO} layer $l$ and \acp{PRB} $k$, respectively. 
Lastly, $\sigma^2_k$ represents the noise power of \acp{PRB} $k$.
Without loss of generality, cells uniformly split their transmit power across \ac{PRB} and beams. 
The effective \ac{SINR} $\gamma_{u,l}$ of \ac{UE} $u$ on \ac{MIMO} layer $l$ is derived from the \acp{SINR} $\gamma_{u,l,k}$ of its allocated \acp{PRB} $k \in {\cal K}{_u}$, following a mutual information framework \cite{1651855}.

\subsubsection*{Achievable UE Rate}

Assuming round-robin scheduling and full-buffer traffic, 
the achievable data rate for \ac{UE} $u$ is calculated as:
\begin{equation}
\label{eq:AchievableDataRate}
    R_u = \sum_{l=1}^{L} \frac
    {N^{\rm PRB}_{\hat{c}_u} \; B^{\rm PRB}_{\hat{c}_u}}
    {N^{\rm UE}_{\hat{c}_u, l}}
    \log_2(1 + \gamma_{u,l}),
\end{equation}
where 
$N^{\rm PRB}_{\hat{c}_u}$ is the total number of \acp{PRB} in the serving cell $\hat{c}_u$, each with bandwidth $B^{\rm PRB}_{\hat{c}_u}$,
and $N^{\rm UE}_{\hat{c}_u, l}$ is the number of \ac{UE} associated with the serving \ac{MIMO} layer $l$.

\subsubsection*{Power Consumption}

For power consumption, 
we employ the realistic model for single and multicarrier \acp{BS} proposed in \cite{piovesan2022machine}, defined as:
\begin{equation}
    \begin{split}
    P_{\mathrm{BS}} &= P_{\rm BBU}  + P_0 + P_{\rm BB} + \\ &+
    \underbrace{ M^{\rm{TRX}}_{\rm{av}} D_{\rm{TRX}}}_{P_{\mathrm{TRX}}} + \underbrace{M^{\rm{PA}}_{\rm{ac}} D_{\rm{PA}}}_{P_{\mathrm{PA}}}  +  \underbrace{\frac{1}{ \eta}  \sum_{c=1}^{C}  P_{\mathrm{TX},c}}_{P_{\mathrm{out}}}, 
    \end{split}
\end{equation}
where \(P_{\rm BBU}\) is the power consumption of the \ac{BBU},
\(P_0\) represents the power consumed by radio unit interfaces and controllers, and \(P_{\rm BB}\) accounts for the baseband processing at the radio unit. 
The term \(P_{\mathrm{TRX}}\) is the power consumed by the radio transceivers, calculated as the product of the number of \acp{TRX} \(M^{\rm{TRX}}_{\rm{av}}\) and the power consumed by each \ac{TRX} \(D_{\rm{TRX}}\).
Similarly, \(P_{\mathrm{PA}}\) is the static power consumed by the \acp{MCPA}, determined by the number of active \acp{TRX} \(M^{\rm{PA}}_{\rm{ac}}\) and the static power consumption per \ac{MCPA} \(D_{\rm{PA}}\).
Note that each \ac{TRX} has its own dedicated \ac{MCPA}, 
and each \ac{MCPA} is responsible for amplifying signals across the multiple carriers used by all cells managed by the radio.
Finally, \(P_{\mathrm{out}}\) represents the total power required to generate the transmit power for all $C$ cells managed by the radio, computed as the total transmit power divided by the efficiency \(\eta\) of the \acp{MCPA} and antennas.  

%
%
\section{Capacity and Power Consumption Analysis}
\label{sec:results}

In this section, we present key results from our analysis of the multi-layer 4G/\ac{5G}/\ac{6G} deployment models discussed earlier. 
The experiments were conducted using Giulia, a system-level simulation tool calibrated with \ac{3GPP} models to ensure accurate and realistic performance evaluation. 

\subsection{Capacity of Multi-Layer 6G Deployments}

\subsubsection*{Standalone 4G vs. 5G}

\begin{figure}[t]
    \centering
    \includegraphics[width=0.45\textwidth]{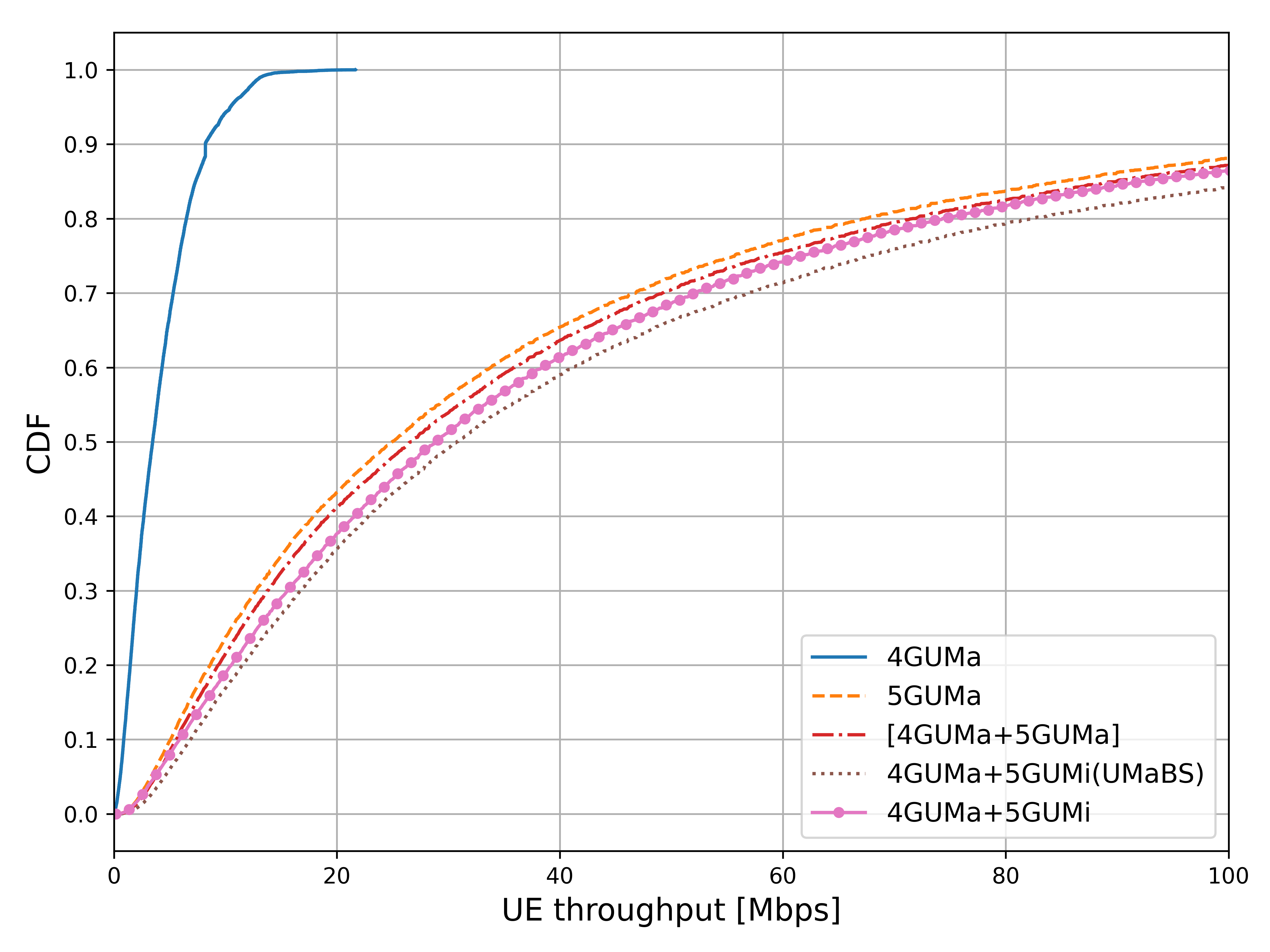}
    \caption{Downlink UE rates (no 6G deployments included).}
    \label{fig:UE_rate_up_to_5G}
\end{figure}

First, consider the \ac{UE} downlink rates in Fig.~\ref{fig:UE_rate_up_to_5G}. 
As expected, \texttt{4G\,UMa} delivers the lowest performance due to fewer cells (57), limited spectrum (20\,MHz), and reduced spatial multiplexing (8TRX). 
In contrast, \texttt{5G\,UMa} performs significantly better, 
with a median \ac{UE} rate of 24.95\,Mbps (a 7.21$\times$ increase) thanks to 5$\times$ more spectrum (100\,MHz) and 8$\times$ higher multiplexing (64TRX) via \ac{CSI-RS} beams. 
The gain, however, is not 40$\times$, as not all \ac{CSI-RS} beams are fully utilized---particularly vertical beams, 
generated using a 2D-DFT codebook and shown in Fig.~\ref{fig:beams}.

\subsubsection*{Co-located 4G and 5G}

The co-located \texttt{[4G\,UMa + 5G\,UMa]} deployment, with more cells (114), improves the median \ac{UE} rate by 6.68\,\% compared to standalone \texttt{5G\,UMa}, as offloading \acp{UE} from 5G to 4G frees up bandwidth per \ac{UE}. However, gains are limited due to relatively small additional bandwidth. Optimized cell reselection thresholds per cell, and not per layer, could also bring further gains. 

\subsubsection*{Non-co-located 4G and 5G}

Non-co-locating 5G with 4G in the \texttt{4G\,UMa + 5G\,UMi} scenario improves the median rate by an additional 15.47\,\% over \texttt{5G\,UMa}. Placing \ac{5G} \acp{BS} closer to high-\ac{UE}-density areas (with a 200\,m \ac{ISD} instead of 500\,m) enhances performance. 
Despite the lower transmit power (44\,dBm vs. 49\,dBm), the proximity of \ac{5G} cells to \acp{UE} drives better results, demonstrating that strategic placement is more effective than merely increasing transmit power.
Increasing the transmit power of the non-collocated \ac{5G} cells to 49\,dBm, as in the \texttt{4G\,UMa + 5G\,UMi (UMa BS)} scenario, further boosts performance by 23.29\,\% over \texttt{5G\,UMa}.

\begin{figure}[t]
    \centering
    \includegraphics[width=0.45\textwidth]{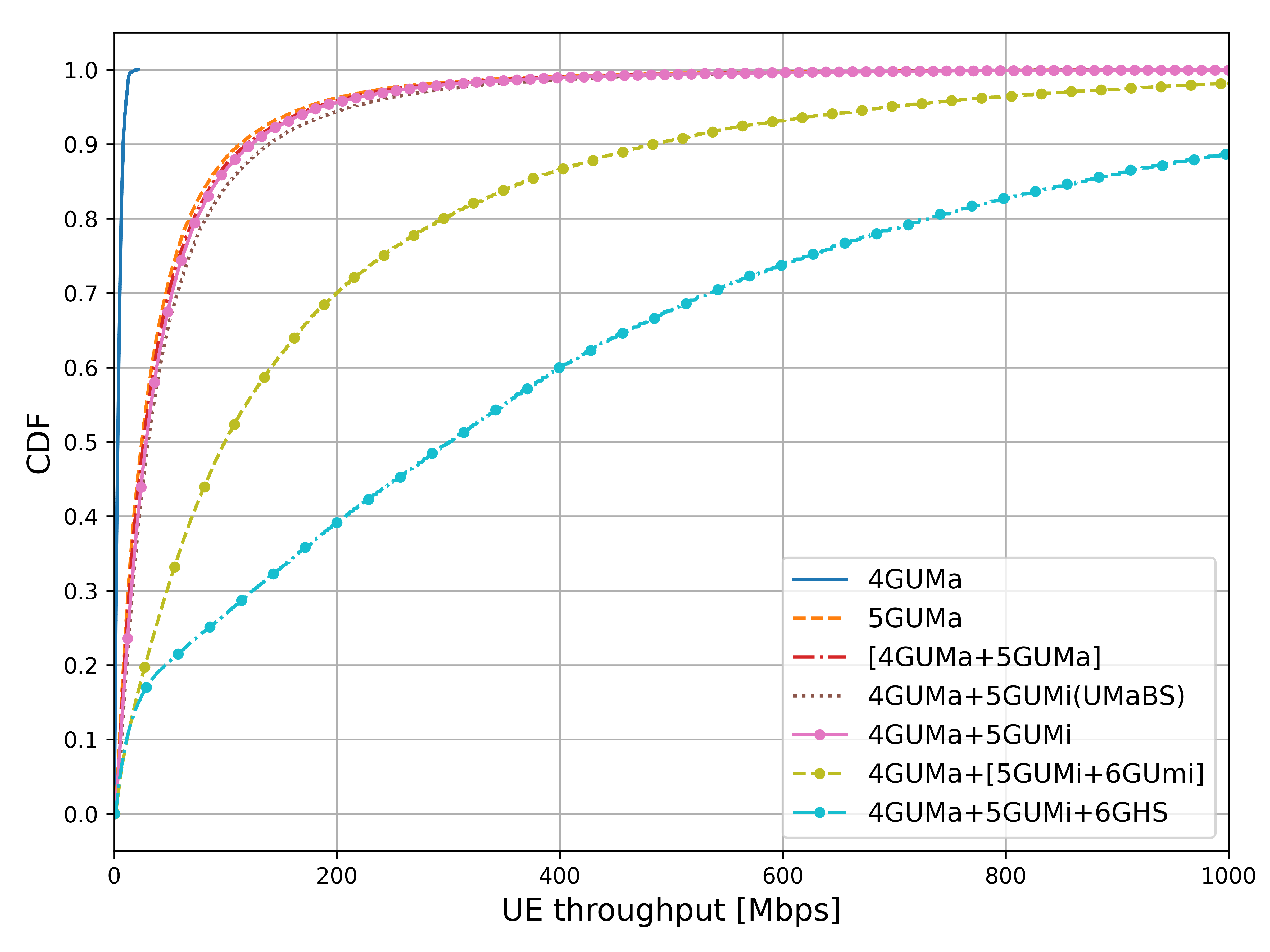}
    \caption{Downlink UE rates (6G deployments included).}
    \label{fig:UE_rate}
\end{figure}

Next, we examine the impact of the 6G \ac{FR3} layer.
Fig.~\ref{fig:UE_rate} extends Fig.~\ref{fig:UE_rate_up_to_5G},
incorporating two 6G deployment scenarios.

\subsubsection*{Deploying 6G on Existing 5G Sites}

In the \texttt{4G\,UMa + [5G\,UMi + 6G\,UMi]} scenario, 
co-locating 6G with 5G introduces 57 new cells, each with 200\,MHz of bandwidth and 128 \ac{CSI-RS} beams, quadrupling the capacity of 5G cells. 
This setup results in a median \ac{UE} rate of 99.64\,Mbps---roughly 4$\times$ higher than the \texttt{4G\,UMa + 5G\,UMi} scenario. 

\subsubsection*{Deploying 6G on New Sites}

The \texttt{4G\,UMa + 5G\,UMi + 6G\,HS} scenario, where 6G pico cells are deployed in hotspot centers instead of 6G micro cells co-located with 5G, leads to significant improvements. This setup delivers a median \ac{UE} rate of 301.06\,Mbps, with peaks exceeding 1\,Gbps (95\%-\emph{tile} \ac{UE} rate of 1.36\,Gbps), even considering the high density of \acp{UE}. 
Despite employing a lower transmit power for 6G cells, this approach triples the median rate of the co-located scenario, highlighting the importance of targeted 6G placement for maximizing network efficiency and throughput.

\subsubsection*{Effect of Cell Reselection}

\begin{figure}[t]
    \centering
    \includegraphics[width=0.45\textwidth]{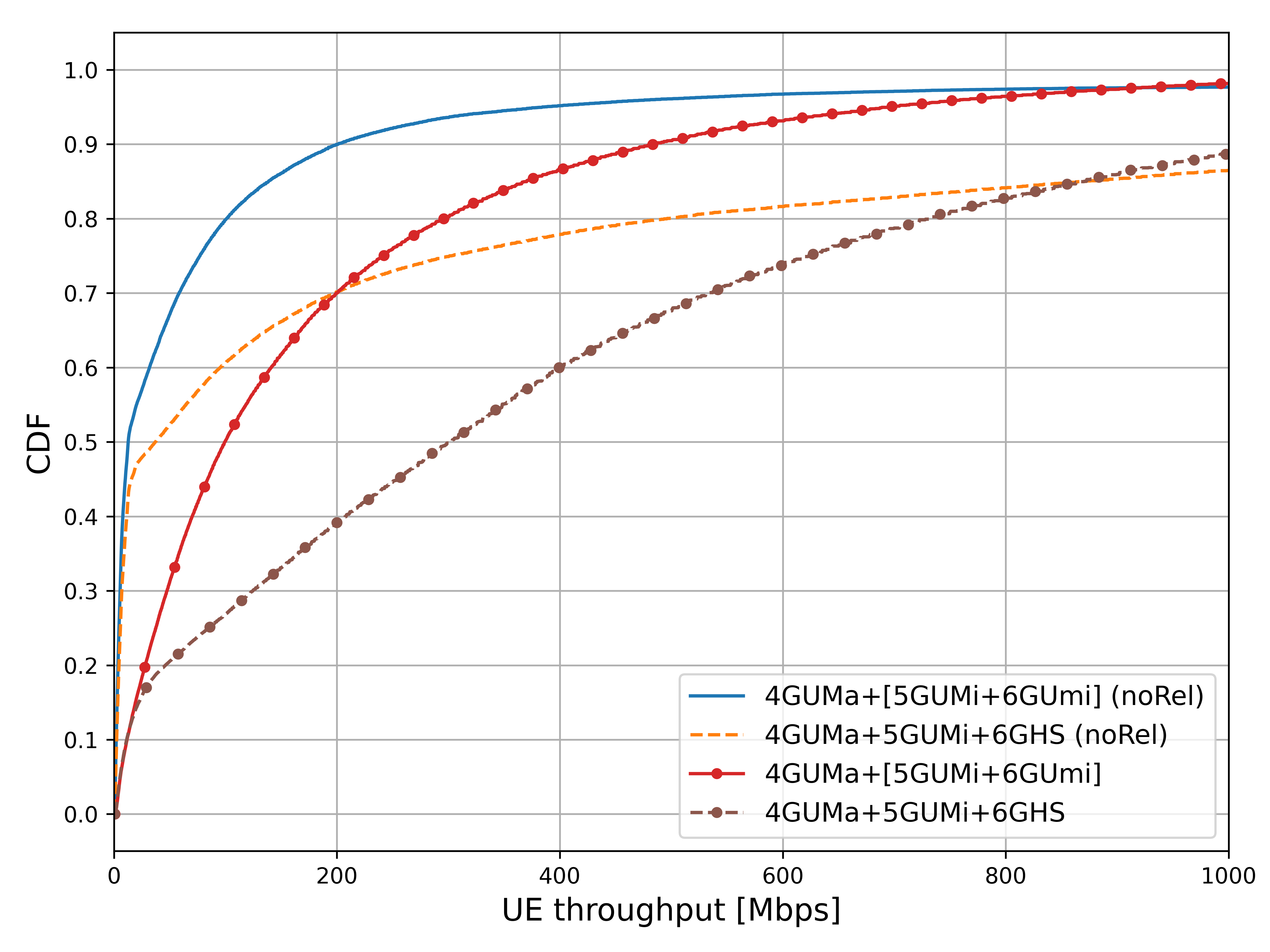}
    \caption{Downlink UE rates with and without cell reselection, where `noRel' denotes no cell reselection.}
    \label{fig:UE_rate_cell_reselction_com}
\end{figure}

Fig.~\ref{fig:UE_rate_cell_reselction_com} shows the impact of priority-based cell reselection on 6G performance. When this procedure is deactivated, forcing \acp{UE} to connect to the cell with the highest \ac{RSRP}, the 6G spectrum remains underutilized, as more \acp{UE} associate with 4G and 5G cells operating at lower frequencies and higher transmit power. This leads to a substantial reduction in network performance, as can be observed in the figure. Optimizing cell reselection is therefore critical to fully exploiting the capacity of multi-layer 6G networks, as important as the strategic placement of 6G cells.

\subsection{Power Consumption of Multi-Layer 6G Deployments}

\begin{figure}[t]
    \centering
    \includegraphics[width=0.45\textwidth]{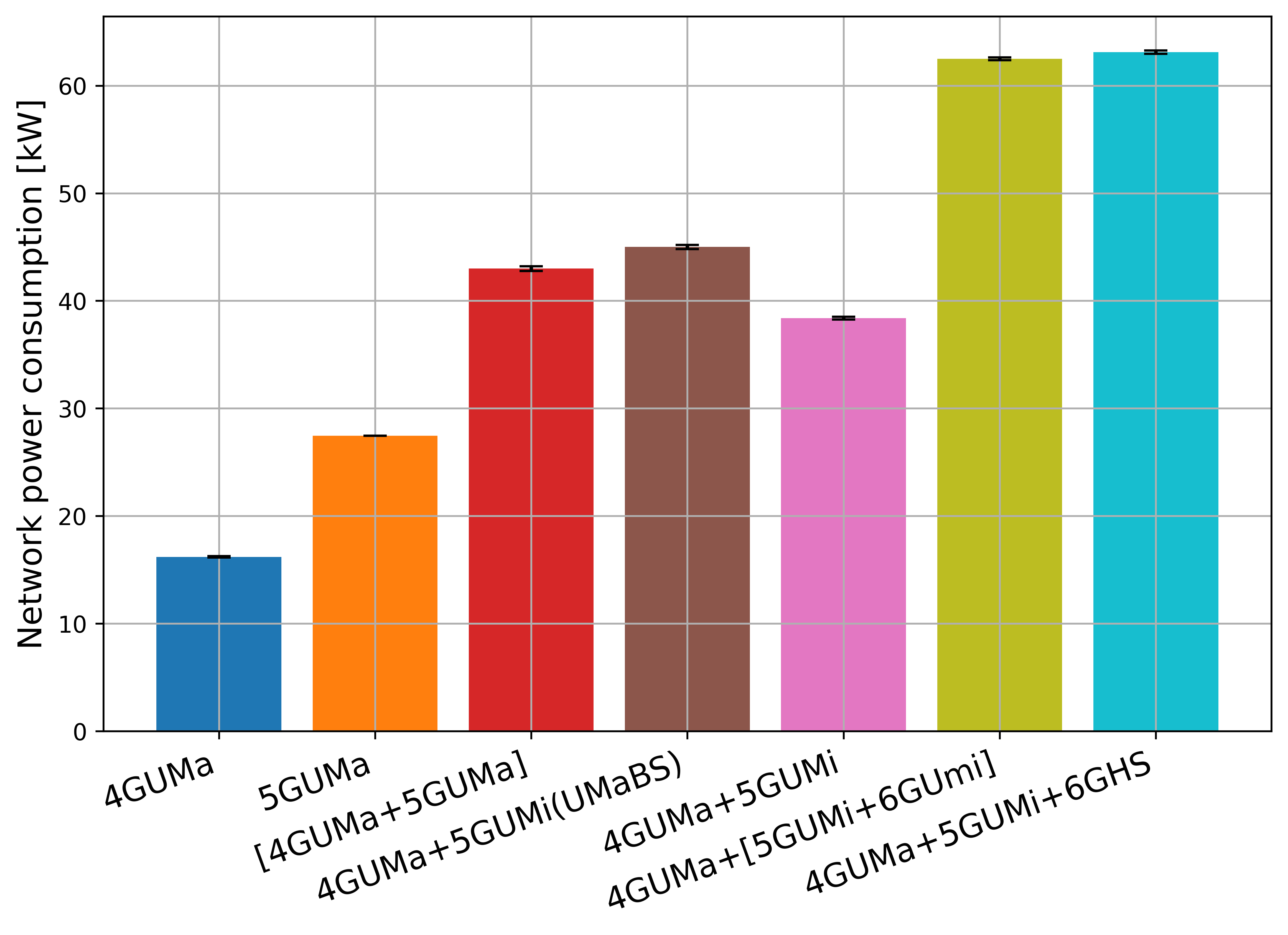}
    \caption{Power consumption under different multi-layer deployments.}
    \label{fig:network_power_consumption}
\end{figure}

Deploying multi-layer 6G networks enhances \ac{UE} performance but comes with increased energy and economic costs.
Fig.~\ref{fig:network_power_consumption} illustrates the significant rise in power consumption that accompanies 5G adoption. 
Co-locating 5G with 4G in the \texttt{[4G\,UMa + 5G\,UMa]} deployment results in a threefold increase in power consumption, 
consistent with real-world deployment observations~\cite{Huawei2020}.

In the non-co-located scenario, \texttt{4G\,UMa + 5G\,UMi (UMa BS)}, 
deploying 5G closer to \acp{UE} results in even higher power consumption due to the additional 5G macro radios. 
However, using smaller 5G micro radios with lower transmit power (44\,dBm vs. 49\,dBm) reduces power consumption compared to the co-located setup, underscoring the importance of optimizing transmit power. 

For the co-located 6G deployment, \texttt{4G\,UMa + [5G\,UMi + 6G\,UMi]}, power consumption increases by 33\,\% compared to \texttt{[4G\,UMa + 5G\,UMa]}. Notably, the non-co-located 6G scenario, \texttt{4G\,UMa + 5G\,UMi + 6G\,HS}, shows that the extra power consumption from new non-co-located 6G sites is modest, thanks to their lower static power requirements. Strategically deploying 6G is crucial, as it maximizes performance gains while keeping the additional power consumption manageable. 
While co-locating with existing 5G sites may reduce deployment costs, it may not always be the optimal strategy for 6G, considering the potential performance benefits of new site placements.

\section{Conclusion}
\label{sec:conclusions}

In this paper, we presented a comprehensive system model for evaluating the capacity and power consumption of multi-layer 6G deployments using the FR3 band. 
Our analysis revealed several key findings with significant implications for future 6G infrastructure deployment:
\begin{itemize}
\item 
Strategic, non-co-located 6G \acp{BS} showed superior performance, achieving median \ac{UE} rates of 300\,Mbps and peak rates exceeding 1\,Gbps, significantly outperforming co-located 5G-6G setups that rely on higher transmit power.
\item
Priority-based cell reselection mechanisms are essential for fully leveraging the capabilities of 6G. Additionally, efficient \ac{CSI-RS} beam configuration will play a critical role in enhancing network coverage and throughput, as many beams may otherwise remain underutilized.
\item 
Although 6G deployment leads to a 33\% increase in power consumption, non-co-located strategies offer a trade-off between performance improvements and energy efficiency.
\end{itemize}
Overall, these findings highlight the importance of a targeted, strategic approach to 6G deployment, optimizing both data rates and power consumption to support sustainable next-generation mobile networks.


\begin{acronym}[AAAAAAAAA]
    \acro{2D-DFT}{two dimensional discrete Fourier transform}
    \acro{3GPP}{3rd Generation Partnership Project}
    \acro{5G}{fifth generation}
    \acro{6G}{sixth generation}
    \acro{AF}{array factor}
    \acro{AH}{aerial highway}
    \acro{AI}{artificial intelligence}
    \acro{AoA}{angle of arrival}
    \acro{AoD}{angle of departure}
    \acro{BBU}{baseband unit}
    \acro{BO}{bayesian optimization}
    \acro{BS}{base station}
    \acro{BVLoS}{beyond visual line of sight}
    \acro{CAGR}{compound annual growth rate}
    \acro{CCUAV}{cellular connected unmanned aerial vehicle}
    \acro{CDF}{cumulative distribution function}
    \acro{cm-wave}{centimeter wave}
    \acro{CQI}{channel quality indicator}
    \acro{CRS}{common reference signal}
    \acro{CSI}{channel state information}
    \acro{CSI-RS}{channel state information-reference signal}
    \acro{D2D}{device to device}
    \acro{DFT}{discrete Fourier transform}
    \acro{DL}{downlink}
    \acro{DoF}{degree of freedom}
    \acro{eGA}{elite genetic algorithm}
    \acro{eICIC}{enhanced inter-cell interference coordination}
    \acro{E}{eastern}
    \acro{ES}{Eigenscore}
    \acro{FR1}{frequency range 1}
    \acro{FR2}{frequency range 2}
    \acro{FR3}{frequency range 3}
    \acro{GA}{genetic algorithm}
    \acro{gUE}{ground user equipment}
    \acro{HO}{handover}
    \acro{ICC}{international conference on communications}
    \acro{IMT}{international mobile telecommunication system}
    \acro{ISD}{inter-site distance}
    \acro{IUD}{inter-UAV distance}
    \acro{ITU}{International Telecommunication Union}
    \acro{LEO}{low Earth orbit}
    \acro{LoS}{line of sight}
    \acro{LTE}{long term evolution}
    \acro{MAMA}{mMIMO-Aerial-Metric-Association}
    \acro{MCPA}{multicarrier power amplifier}
    \acro{MINP}{mixed-integer nonlinear problem}
    \acro{ML}{machine learning}
    \acro{MIMO}{multiple-input multiple-output}
    \acro{mMIMO}{massive multiple-input multiple-output}
    \acro{mmWave}{millimeter wave}
    \acro{MNO}{mobile network operator}
    \acro{MU-mMIMO}{multi-user massive multiple-input multiple-output}
    \acro{MU-MIMO}{multi-user multiple-input multiple-output}
    \acro{NLoS}{non line of sight}
    \acro{NOMA}{non-orthogonal multiple access}
    \acro{NR}{new radio}
    \acro{PBCH}{physical broadcast channel}
    \acro{P2P}{point to point}
    \acro{PAHSS}{Particle Aerial Highway Swarm Segmentation}
    \acro{PL}{path loss}
    \acro{PMI}{precoding matrix indicator}
    \acro{PRB}{physical resource block}
    \acro{PSO}{particle swarm optimization}
    \acro{PSS}{primary synchronization signal}
    \acro{QoS}{quality of services}
    \acro{RAN}{radio access network}
    \acro{RE}{resource element}
    \acro{RI}{rank indicator}
    \acro{RRC}{radio resource control}
    \acro{RSRP}{reference signal received power}
    \acro{RSS}{received signal strength}
    \acro{SRS}{sounding reference signal}
    \acro{SSB}{synchronization signal block}
    \acro{SINR}{signal-to-interference-plus-noise ratio}
    \acro{SO}{southern}
    \acro{SSS}{secondary synchronization signal}
    \acro{SVD}{single value decomposition}
    \acro{thp}{throughput}
    \acro{TRX}{transceiver}
    \acro{UAM}{urban air mobility}
    \acro{UAV}{unmanned aerial vehicle}
    \acro{UE}{user equipment}
    \acro{UL}{uplink}
    \acro{UMa}{urban macro}
    \acro{UMi}{urban micro}
    \acro{UPA}{uniform planar array}
    \acro{UPi}{urban pico}
    \acro{URD}{Urban Random Distributed}
    \acro{URLLC}{ultra-reliable low latency communication}
    \acro{WRC}{world radiocommunication conference}
    \acro{ZF}{zero forcing}

\end{acronym}

\bibliographystyle{IEEEtran}
\bibliography{journalAbbreviations, main}

\end{document}